\documentclass{rspublic}
\usepackage{psfig}

\def\mnras{MNRAS}
\def\apj{ApJ}\def\apjl{ApJ}
\def\etal{{\it et al.}\,}
\def\msun{\,{\rm M}_\odot}

\begin{document}

\title[Cooling and Clusters]{Cooling and Clusters: When Is Heating Needed?}

\author[G. Bryan and M. Voit]{Greg Bryan$^1$ and Mark Voit$^2$}

\affiliation{$^1$ Astrophysics, Oxford University, Oxford OX1 3RH\\
$^2$Dept. of Physics and Astronomy, Michigan State University, East
Lansing, MI }

\label{firstpage}

\maketitle

\begin{abstract}{galaxies: clusters -- galaxies: cooling flows --
    galaxies: formation -- galaxies: active -- methods: numerical }
There are (at least) two unsolved problems concerning the current
state of the thermal gas in clusters of galaxies.  The first is
identifying the source of the heating which offsets cooling in the
centers of clusters with short cooling times (the ``cooling flow''
problem).  The second is understanding the mechanism which boosts the
entropy in cluster and group gas.  Since both of these problems
involve an unknown source of heating it is tempting to identify them
with the same process, particular since AGN heating is observed to be
operating at some level in a sample of well-observed ``cooling flow''
clusters.  Here we show, using numerical simulations of cluster
formation, that much of the gas ending up in clusters cools at high
redshift and so the heating is also needed at high-redshift, well
before the cluster forms.  This indicates that the same process
operating to solve the cooling flow problem may not also resolve the
cluster entropy problem.

\end{abstract}

\section{Introduction}

The physical state of the gas in clusters of galaxies tells us a great
deal about the past history of the cluster; however, it is not simple
to decode this information.  During a typical 10 Mpc (comoving) trip
from its initial location at early times to its current resting place,
the gas may be heated by shocks, galactic winds, AGN, turbulence and
thermal conduction; it may also be cooled by radiative cooling.  It is
worth noting that of all of these processes, only radiative cooling
will always act to reduce specific entropy of the gas, while the others
(usually) act to boost the gas to a higher adiabat.  It is perhaps not
surprising then to find that observations of clusters indicate the gas
is at a higher entropy than found in simulations that include only
gravitational infall and shocks (Ponman \etal 1999).  Moreover, this
effect is not uniform across all clusters -- the entropy boost is
relatively larger for small clusters, which have lower intrinsic
entropy, than for larger clusters.

It is this relative effect which changes the slope of the relation
between X-ray luminosity and gas temperature in clusters from the
self-similar prediction of $L_X \sim T^2$ to the observed $L_X \sim
T^3$ (Edge \& Stewart 1991; Evrard \& Henry 1991; Kaiser 1991).
Clearly whatever process raised the entropy, it was operating more
effectively for groups than for clusters.  However, this alone does
not determine the source of the heating.  On energetic
grounds, none of the sources listed earlier can be excluded
(e.g. Finoguenov \etal 2000; Tozzi \& Norman 2001; Cavaliere \etal
2002; Wu \etal 2001; Kim \& Narayan 2003).

The other piece of evidence for heating in clusters comes from the
absence of cooling in so-called ``cooling flow'' clusters (e.g. Fabian
2003).  Probably the most promising resolution to the cooling-flow
riddle comes from AGN heating.  It is now well established from
Chandra and XMM-Newton observations that in the centers of many
clusters with cooling times short compared to the Hubble time, there
are cavities thought to be inflated by jets powered by a supermassive
black hole (Fabian \etal 2000; Mazzotta \etal 2002; McNamara \etal
2001).  While the exact mechanism for transferring energy to the gas
is not yet perfectly clear there is no shortage of work on this topic
(Churazov \etal 2001; Br{\"u}gen \& Kaiser 2001, 2002; Omma \etal 2004;
Reynold, \etal 2004; Begelman \& Ruszkowski, these proceedings).

What is not clear is if these two pieces of evidence are related.  In
other words, is it the same source of heating which solves the cooling
flow problem AND reproduces the correct entropy distribution (and hence
thermal structure) of X-ray clusters?

In this work, we will first examine a simple model based on the
characteristics of cooling and then use numerical simulations to find
out at which epoch most of the cooling (and hence heating) occurs.  If
we know when the heating must occur, this may cast some light on the
source of the heating.  We will show that the epoch of cooling is
generally at much higher redshift than the formation of clusters and
so the solution to the $z=0$ cooling flow problem may not also be the
solution to the structure problem.

\section{Understanding cluster structure}

One early model for understanding what appeared to be a floor in the
entropy distribution was that the gas was uniformly heated to a high
constant adiabat at early times (e.g. Evrard \& Henry 1991; Kaiser
1991).  While this model did correctly reproduce the observed
luminosity-temperature relation (Bialek et al. 2001), it suffered from
possible conflicts with the observed low entropy of the Lyman-alpha
forest.  In addition, the isentropic cores predicted by this model for
groups have not been observed (e.g. Pratt \& Arnaud 2003; Ponman \etal
2003).

There have been a number of suggested models which attempt to
reproduce the structure and scaling of clusters based on particular
ways of adding energy to the cluster gas (e.g. Wu \etal 2001; Bower
\etal 2001), but it is clear that the result depends on when and where
the heating occurs.  Instead, here we will focus on a simple model
which is based on the entropy distribution and doesn't specify exactly
how the heating occurs.  This approach is a useful framework for
understanding what physics is required to correctly reproduce the
cluster structure and hence reproduce the cluster scaling relations.

To understand these scaling properties -- such as the $L_X-T$ relation
-- it is useful to create an idealized, hydrostatic, spherical model
of a cluster in a fixed dark matter potential based on N-body
simulation.  Then, given an appropriate boundary condition, the state
of the gas is entirely specified by the specific entropy distribution.
This is true because, in equilibrium, the entropy must be a
monotonically increasing function of radius (otherwise convection will
occur).

\begin{figure}
\centerline{\psfig{file=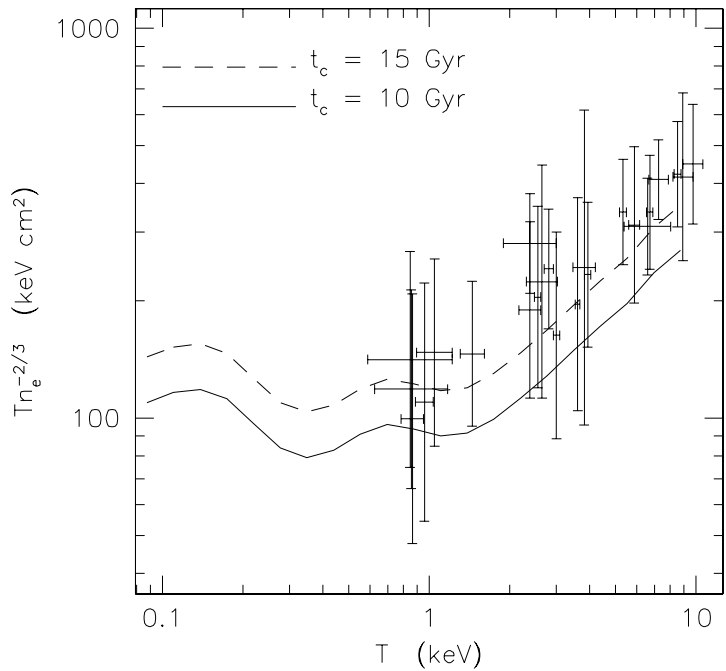,width=.75\hsize}}
\caption{This plot shows a measure of the entropy measured at 10\% of
the virial radius against cluster gas temperature for a collection of
clusters from Ponman et al. 1999.  Measuring at this radius makes the
results insensitive to any central cooling cusp.  The points have a
slope which is much flatter than the self-similar $K = T n_e^{-2/3}
\sim T$.  The solid and dashed lines show the locus of points where
the cooling time is equal to the value indicated.}
\label{fig:entropy}
\end{figure}

It has become usual to define the mass $M$ of a cluster as the mass
within a fixed overdensity, so that the characteristic density $\rho
\sim M/R^3$ of all clusters at a given epoch is a constant times the
critical density of the universe (see e.g. Bryan \& Norman 1998).  The
temperature then scales as $T \sim M/R$.  We can define a measure of
the entropy for an ionized monatomic ideal gas, $K = T n_e^{-2/3}$,
where $n_e$ is the electron density corresponding to a completely
ionized gas at density $\rho$.  This is a useful definition because it
can be measured observationally (e.g. Ponman, Sanderson and Finoguenov
2003) and the results are shown in figure~\ref{fig:entropy} for a
collection of clusters.  If we use the self-similar model, then $K$
should scale simply as $K \sim T$ (since we have assumed that the
density --- and therefore $n_e$ --- is the same for all clusters at a
given redshift).

\begin{figure}
\centerline{\psfig{file=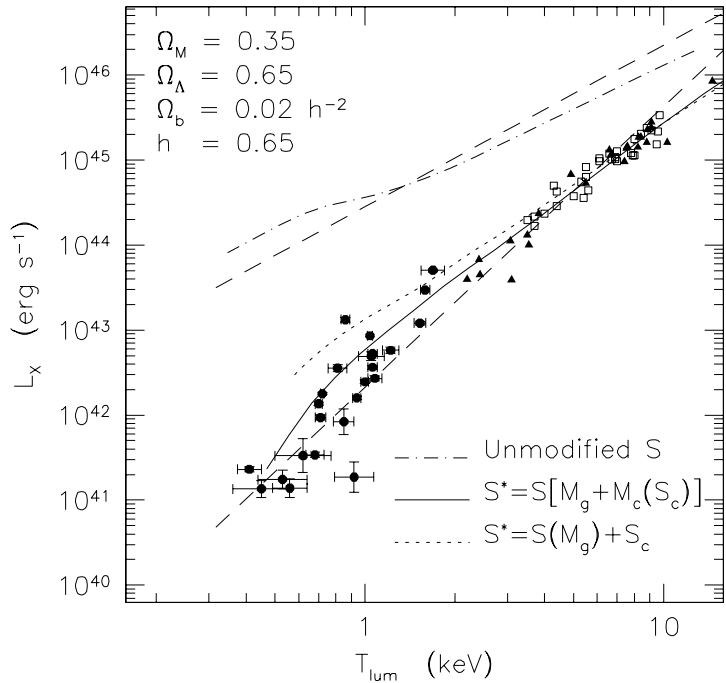,width=.75\hsize}}
\caption{The relation between bolometric X-ray luminosity and
  luminosity-weighted temperature for samples of observed clusters
  from Arnaud and Evrard (1999), Markevitch (1998) and Helsdon \&
  Ponman (2000).  The dot-dashed line is the predicted relation from a
  hydrostatic equilibrium model with an unmodified entropy
  distribution, while the solid and dotted lines are the same model
  with low entropy material removed or heated, respectively.  The assumed
  cosmological parameters for these models are shown in the upper left.
  The lower (upper) dashed line is a fit to numerical
  simulation with (without) radiative cooling from Pierce \etal (2000).
}
\label{fig:lt}
\end{figure}

Clearly the results do not agree with the self-similar prediction ($K
\sim T$), but do match the locus of points where the cooling time
equals the age of the universe (a reasonable stand-in for the age of
the cluster).  This is consistent with a model in which all the gas
that is below this line has cooled and either formed stars or been
re-heated by supernovae or AGN.  This idea was explored further by
Voit \& Bryan (2001) and Wu \& Xue (2002) who constructed a simple
spherically symmetric, hydrostatic model that used the entropy
distribution from simulations without cooling, star formation or
feedback.  As shown by the dash-dot line in figure~\ref{fig:lt}, the
resulting $L_X-T$ relation agrees with simulations in which there is
no cooling (upper dashed line), but it does not agree with
observations.  When they excluded the low entropy gas that had a
cooling time below the Hubble time (i.e. below the line in
figure~\ref{fig:entropy}), either by simply removing it, or by
shifting the entropy distribution so that no gas was below the
critical ``cooling'' entropy, then the resulting $L_X-T$ relation
matched the observations (and simulations that include cooling), as
can be seen by the lower lines in figure~\ref{fig:lt}.

This model does not tell us how the gas was heated, it simply predicts
how much of the gas needs to have its entropy modified.  In fact, it
works equally well if the low entropy gas is simply removed and turned
into stars (Bryan 2000); however, in this case the resulting baryon
fraction in stars exceeds the observed value (e.g. Balogh \etal 2001).

Although the model reproduces observations it is open to a number of
criticisms.  One is that it implicitly assumes that the fraction of
gas that cools and may be re-heated can be found from the present-day
distribution of matter.  In hierarchical cluster formation models like the cold dark
matter (CDM) one, the cluster is formed out of smaller objects at high
redshift, and it is not clear if the fraction at $z=0$ is
representative of the total fraction that would cool over the
cluster's lifetime.

In order to understand the evolution of the gas in clusters better, a
number of approaches are possible.  One is the construction of
analytic models of accretion and shock-heating.  For example, these
are explored extensively in Voit, Bryan \& Balogh (2002) and Voit
\etal (2003).  Here, on the other hand, we examine what can be learnt
about the build-up of clusters through numerical simulations.

\section{Insight from simulations}

We examine a simulation of a typical massive cluster in a
cosmological-constant dominated, spatially flat CDM model with a
Hubble constant of 70 km s$^{-1}$ Mpc$^{-1}$ and $\Omega=0.3$ (the
ratio of the matter density to the critical density).  The simulation
was performed with an adaptive-mesh refinement (AMR) technique (Bryan
1999).  AMR is a grid-based hydrodynamics method that starts with a
uniform mesh and adds additional, finer grids as required to model the
collapsing structures.  Its strengths are that it can both model
shocks well, and provide high spatial resolution in regions of
interest.  Dark matter is modelled through collisionless particles that
interact only via gravity, which is computed with Poisson's equation
using an adaptive particle-mesh technique (O'Shea \etal 2004).

The cluster we examine has a mass of $7 \times 10^{14}\msun$, where
this virial mass is defined as the mass within a sphere that has a
mean density 200 times the critical density.  The luminosity-weighted
temperature is 5 keV.  The dark matter particle mass in the simulation
is $m_{\rm dm} = 1.6 \times 10^9 \msun$ and there are about 400,000
particles in the virial radius at the final epoch.  The highest
resolution resolved by the adaptive mesh is 10 kpc. Radiative cooling is
turned off.

\begin{figure}
\centerline{\psfig{file=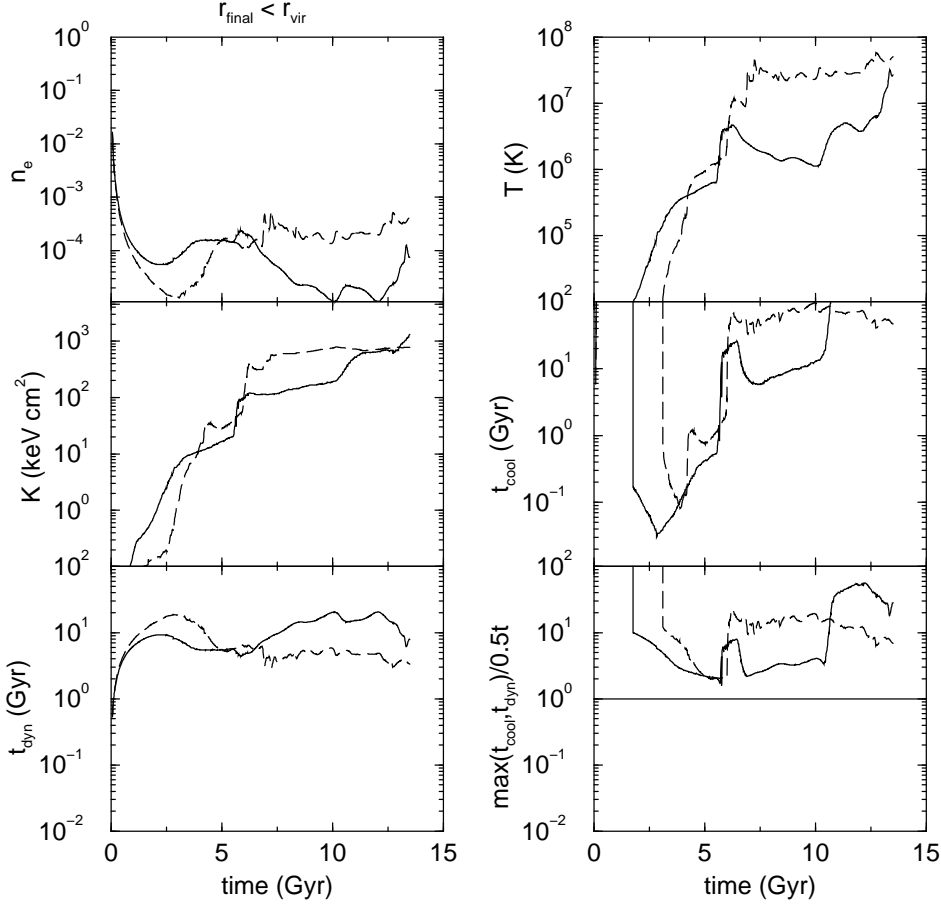,width=.99\hsize}}
\caption{These panels show the physical conditions that a massless
  test particle encounters as it moves with the gas flow during the
  formation and evolution of a typical massive cluster.  Two flow
  lines are drawn at random from those that end up within the virial
  radius of the cluster and shown in each graph.  The electron density
  (upper left) is in units of cm$^{-3}$.  The middle panel on the left
  is the same measure of entropy as shown in figure~\ref{fig:entropy}}
\label{fig:trace1.0}
\end{figure}

In order to study the evolution of the gas which ends up in the
cluster, we have developed a form of massless test particle that moves
with the gas flow.  The trajectories of these particles are then
time-dependent flow lines through the forming cluster, and we can
record the changing conditions along these trajectories.  The test
particles are laid down on a uniform grid at high redshift when the
density distribution is nearly uniform so that each one is a
representative sample of fixed mass fraction of the cluster.  We show
in figure~\ref{fig:trace1.0} an example of two such trajectories
randomly selected from those that end up within the virial radius of
the cluster at the final time.

The power-law drop in the density at early times (seen in the plot of
the electron density) comes from the expansion of the universe.  This
is reversed after a few Gyr when the particle feels a significant pull
from nearby substructure.  Based on the relatively small increase in
density seen in this figure, it is likely that the gas associated with
these tracer particles is in some sort of filamentary structure or in
the outskirts of a virialized halo rather than deep inside a halo.

As the trajectories progress in time, they come to be associated with
larger sub-structures and the temperature increases, as does the
entropy.  In particular, there is a major merger shortly after the 5
Gyr mark which appears as a jump in the entropy of nearly all of the
trajectories.  This merger is clearly associated with a strong shock
which propagates from smaller to larger radii and boosts the
entropy of almost all of the gas.  Note also the nearly monotonic
increase in entropy\footnote{We are following a massless test particle
which is not the same thing as a fluid element, so a local decrease in
entropy due to mixing with lower entropy gas is physically possible.}.
This increase in temperature makes, at first, for a decrease in the
cooling time (computed assuming 1/3 solar metallicity).  As the
temperature grows beyond the cooling peak at $10^5$ K, the cooling
time then increases dramatically.

Since the simulation does not include radiative cooling, we 
gauge whether the local gas will cool and
condense into stars as follows. We take the maximum of the cooling time and the
local dynamical time $t_{\rm dyn}=(3\pi/16 G\rho)^{1/2}$ and divide this
by one-half of the age of the universe (at that time).  If this
fraction is less than 1, catastrophic cooling is likely.  Using
this measure, we see from the last panel that cooling is not likely to
be important for the two trajectories shown in figure~\ref{fig:trace1.0}.

\begin{figure}
\centerline{\psfig{file=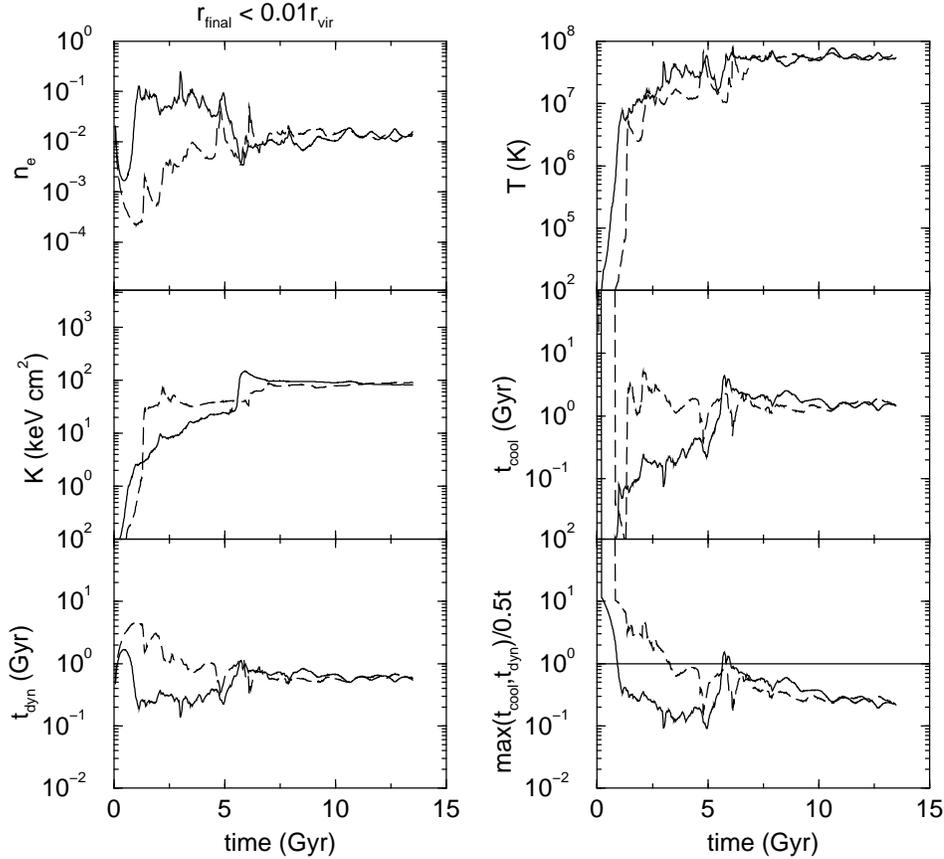,width=.99\hsize}}
\caption{The same as figure~\ref{fig:trace1.0} but for particles
  which end up in the center of the cluster, within 1\% of the virial
  radius.}
\label{fig:trace0.01}
\end{figure}

The situation is quite different for the two trajectories shown in
figure~\ref{fig:trace0.01} which shows physical conditions along two
flow lines which end up in the center of the cluster (more precisely,
within 1\% of the virial radius of the center).  In this case, the
power-law decrease of the density is very quickly reversed as the
particles fall into massive halos that form at high-redshift.  The
associated gas shocks to high temperature and a relatively high
entropy level.  However, unlike in the previous case, the entropy
stays relatively constant after that.  Only one of the two trajectories
experiences the strong shock associated with the merger at $t \sim 6$
Gyr and in this case the density decreases while the temperature is
unchanged.  This is likely to be associated with some sort of
ram-pressure stripping event.  Note also that the density and
temperature values along the trajectories appear to be noisy while the
gas stays on nearly the same adiabat.  This comes from merger-driven
turbulence which adiabatically expands and compresses the gas.

The cooling time histories are also different for these
trajectories which end up in the center of the cluster, with
relatively short cooling and dynamical times throughout.  This is
due mostly to the high densities encountered.  In fact, the later
history should not be taken too seriously because as can be seen in
the lower-right panel of this figure, the two points cross the
critical cooling curve quite early in their history and would be
expected to have either formed stars or have been re-heated by
supernovae or AGN.  

From these two figures we see that gas which ends up in the center of
clusters has quite a different history than the rest of the cluster
gas.  It falls into massive halos very early and will surely have been
involved closely with some star formation event at high-redshift.

\begin{figure}
\centerline{\psfig{file=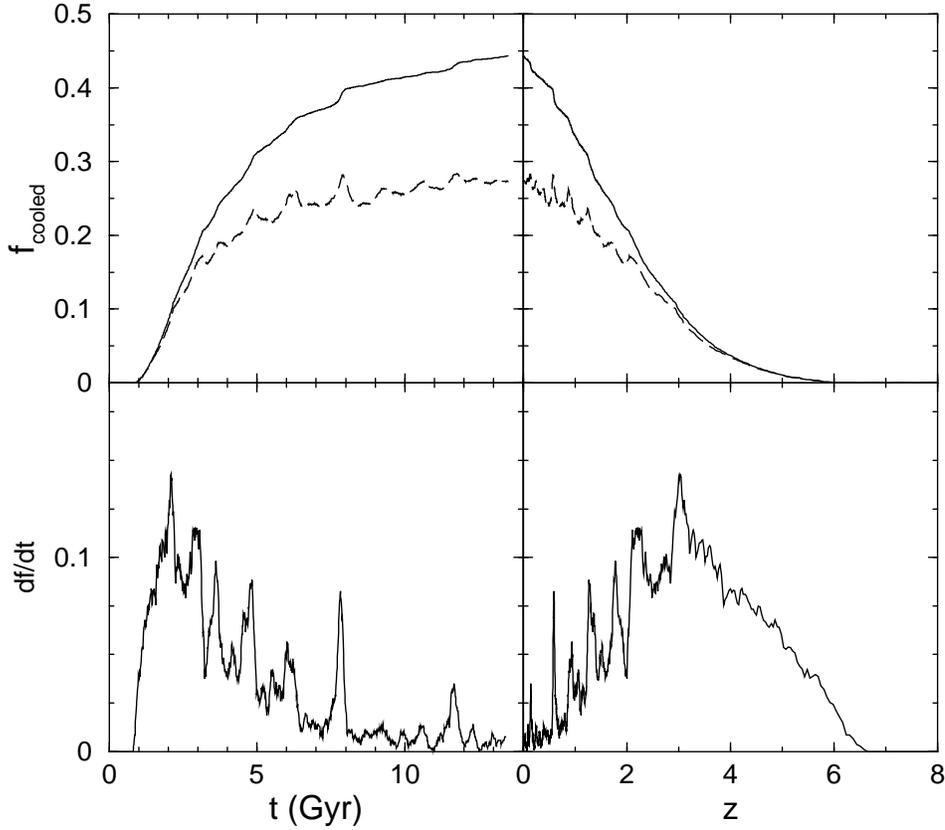,width=.99\hsize}}
\caption{The top two panels show, as a function of time (left) and
  redshift (right), the fraction of trajectories of
  massless test particles (and hence of mass in the cluster) which satisfy
  our joint criteria of a short cooling time and a short dynamical
  time and so are likely to have cooled.  The solid line is the
  fraction of trajectories which have met this criterion at any time in
  the past, while the dashed line shows the instantaneous fraction.
  The bottom two panels show the derivative of the solid curve (which
  is non-negative by definition).}
\label{fig:fcool}
\end{figure}

While it is instructive to examine individual trajectories, we can go
beyond this.  Because the test particles were laid down uniformly at
high redshift they are representative (in a mass-weighted sense) of
the distribution of gas which ends up in the cluster.  Therefore, we
can use their trajectories to create distribution functions of gas
properties (density, temperature, entropy) as a function of time for
gas which will be in the cluster at $z=0$.  As a first step along this
road, we have computed the fraction of trajectories which satisfy the
``cooled'' criterion defined earlier (short cooling and dynamical
times).  In figure~\ref{fig:fcool}, we show, as a solid line, the
fraction of particles that have met this criterion at any time in
their past, and so would have -- in the absence of feedback -- cooled
to form stars.  We also show the instantaneous fraction as a dashed
line for comparison.  While this second value is obviously less
accurate it does provide a roughly similar estimate of the total
cooled fraction and, at $z=0$, corresponds to the fraction of gas
below the critical ``cooling'' entropy calculated in the model in
section 2 (see also Voit \& Bryan 2001).

In the bottom panels of the same figure, we show the differential of
the cooled fraction, which can be interpreted as the cooling rate of
the gas.  This is shown both as a function of time and redshift.
It is interesting to note that there is a burst of cooling a Gyr or
two after the big-bang, followed by a long tail to late times.

\section{Discussion}

Figure~\ref{fig:fcool} goes a long way to answering the question
we posed in the beginning -- when does the cooling occur?  Clearly the
vast majority of cooling (and hence heating if we are not to generate
too many stars) must occur before $z \sim 1$, which is the epoch of
the formation of clusters.  Therefore, it seems unlikely that any
heating process which depends on having a substantial amount of hot
gas can solve both the $z=0$ cooling flow problem and the cluster
structure/overcooling problem.  In particular, processes involving
thermal conduction, turbulence in clusters, or the dissipation of heat
in sound waves are disfavoured as solutions for the cluster structure
problem.

The peak of the cooling rate is at $z \sim 3$ with a long tail to $z
\sim 6$, which overlaps significantly with the epoch of quasar
formation so some other form of AGN jet heating is possible.  However,
it is not obvious how to couple the heating from AGN to the gas at
large redshift.  AGN jets are often observed to expand well beyond the halo of
hot gas associated with elliptical galaxies and mechanical heating
models are problematic if this is the norm.  In addition, if the gas
actually does cool onto massive halos and is then heated, a great deal
of energy is required not only to remove the gas from the potential of
the galaxy but also to heat it to the minimum entropy levels seen in
figure~\ref{fig:entropy}.  One way in which these requirements are
lessened is if feedback works in tandem with gravitational infall and
shocking.  If AGN can effectively smooth the gas distribution, then
Voit \etal (2003) have shown that the resulting accretion onto
clusters generates more entropy than if the accretion was clumpy.


\begin{acknowledgements}
GLB acknowledges support from PPARC grant PPA/G/O/2001/00016 and the
Leverhulme foundation.
\end{acknowledgements}

\label{lastpage}
\end{document}